\documentclass[%
 reprint,
 amsmath,amssymb,
 aps,
]{revtex4-2}

\usepackage{graphicx}
\usepackage{dcolumn}
\usepackage{bm}

\begin{document}

\preprint{APS/123-QED}

\title{Characterizing multielectron dynamics during recollision}

\author{Graham G. Brown \& Dong Hyuk Ko}
\author{Chunmei Zhang}
 \email{chunmei.zhang@uottawa.ca}
\author{P. B. Corkum}
\affiliation{%
 Joint Attosecond Science Laboratory, Department of Physics, University of Ottawa, Ottawa Canada K1N 6N5\\
 National Research Council of Canada, 100 Sussex Drive, Ottawa, Canada K1A 0R6
}%

\date{\today}

\begin{abstract}
Measuring the delay for an electron to emerge from different states is one of the major achievements of attosecond science \cite{01Goulielmakis2010}. This delay can have two origins – the electron wave packet is reshaped during departure by the electrostatic field of the ionizing medium or it is modified by dynamic interaction with the remaining electrons \cite{02PhysRevLett.111.233005}. Most experiments have observed the former, but confirmation requires a complex calculation. A direct measurement of multielectron dynamics is needed. Photo-recombination – the inverse of photoionization – occurs naturally during electron recollision and can be measured by combining a perturbing beam to modify the recollision electron before recombination \cite{03Dudovich2006}. These “\emph{in situ}” methods allow us to unambiguously isolate multielectron dynamics – the reference being the spectral phase of an attosecond pulse simultaneously measured in spectral regions without multielectron interaction. Here, we measure the group delay of the recollision electron caused by plasmonic resonance dynamics in Xe, simulate the \emph{in situ} measured spectral phase of a recollision electron generated in the presence of the plasmonic resonance in C$_{60}$ and present a corresponding semi-classical theory based on the strong-field approximation. Our results suggest that \emph{in situ} techniques, together with 300 eV recollision electrons, will allow the ultimate time response of electronic matter to be measured. 
\end{abstract}

\maketitle

When an electron in a collision experiment approaches a quantum system, the system feels a time-dependent electric field that depends on the electron’s impact parameter and energy \cite{04Schippers_2019}.  This time-dependent field can induce virtual or real transitions in the system which, in turn, emit their own time-dependent electric field. In response, the incident electron’s energy and direction are modified, encoding the multielectron response onto the spatial structure of the beam. Because of their generality, collisions remain the primary way that we probe multielectron dynamics in atoms or molecules.  

The most important challenge for attosecond science is to develop methods to characterize the time response of multielectron dynamics in atoms, molecules, and solids \cite{05Corkum2007}. However, except for the collision physics-like nonsequential ionization \cite{06Li2020}, recollision, which melds collision and optical physics, has largely ignored its implications for studying multielectron effects, even though the three-step model of attosecond pulse generation indicates that the dynamics of the recollision electron is encoded in the spectral properties of the generated attosecond pulses \cite{07PhysRevLett.96.223902}. We will show that measuring the spectral phase of attosecond-pulse-producing electrons as a function of energy will provide access to the full-time response of multielectron systems.  

There are two classes of attosecond pulse measurement methods. The first, known as attosecond streaking, relies on the photoelectric effect creating an electron replica of the attosecond pulse by photoionizing a selected atom and the velocity distribution of the replica being “streaked” by the time dependent field of a phased, co-applied infrared pulse.   For a well-understood photoemitter, attosecond streaking fully characterizes an attosecond pulse \cite{08Paul1689}.  However, a streaking measurement contains information about both the generating and measuring atom, including multielectron dynamics and structural effects. These disparate effects can be difficult to deconvolve.  

The second class of measurement, often called “\emph{in situ}” techniques, measures the continuum time of the recolliding electron \cite{03Dudovich2006, 09PhysRevA.80.033836}. They have been shown to be insensitive to structure that affect the phase of emission but not the continuum dynamics \cite{10PhysRevA.94.023825}. For an \emph{in situ} measurement, the electric field of a weak pulse perturbs the recollision electron trajectories, thereby modifying the spectral or spatial properties of the attosecond pulse. The variation in the generated XUV spectra as a function of the relative phase $\phi$ between the perturbing and driving pulses can be inverted to determine the spectral properties of the electron continuum propagation, including trajectory resolved recombination and ionization times. Both attosecond streaking and \emph{in situ} measurement have parallels with Frequency-Resolved Optical Gating for Complete Reconstruction of Attosecond Bursts (FROG CRAB) \cite{11PhysRevA.71.011401} and ptychography \cite{12Leblanc2016}. 

\begin{figure}
	\includegraphics[width=\columnwidth]{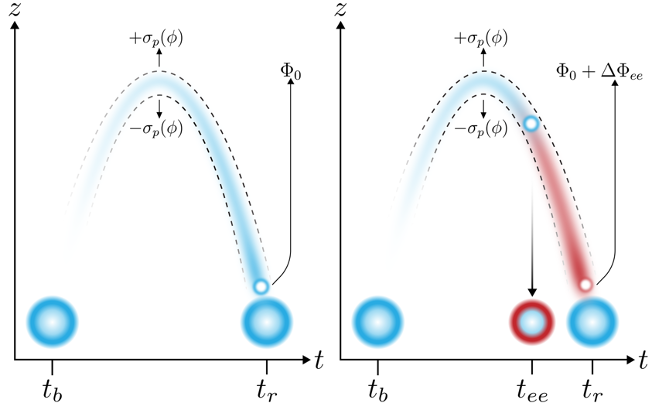}
	\caption{In a correlated system, an electron has two pathways to recombination: the direct (a) and correlated channel (b). In the direct channel, an electron enters the continuum at time $t_b$ with momentum $\vec{p}_0$, is driven by a strong laser field, and recombines at time $t_r$, resulting in the emission of an XUV photon. In the correlated channel, an electron enters the continuum at time $t_b$ with momentum $\vec{p}_0 = \vec{p} - \vec{q}$, interacts and transfers momentum $-\vec{q}$ and energy (depicted by the downward arrow),  to a bound electron at time $t_{ee}$, and then recombines into the ground state at time $t_r$, resulting in the emission of an XUV photon. The change in momentum throughout the trajectory is indicated by the color of the trajectory in (b), going from blue to red after the electron-electron interaction, and the excited ion state is colored similarly. The electron-electron interaction  process modifies the phase of the recolliding electron by an amount $\Delta \Phi_{ee}$. By perturbing the electron trajectories with a weak perturbing pulse, an additional momentum-dependent phase $\sigma_p$ is imparted onto the recolliding electron of momentum $p$, which is opposite in sign for adjacent half-cycles (shown by dashed lines) and results in the emission of even harmonics. The variation of the even-harmonic signal can be used to characterize the phase of the recolliding electron, providing a direct measurement of the electron-electron interaction phase.}	
\end{figure}

Multielectron interactions occurring in the continuum are necessary to explain plasmonic enhancement of the high harmonic spectrum of Xe \cite{02PhysRevLett.111.233005, 13Shiner2011}. These correlated interactions modify the trajectory dynamics and the spectral properties of the emitted attosecond pulse, as sketched in Fig. 1.  Figure 1A depicts direct recollision: the electron enters the continuum at time $t_b$, is accelerated by the driving field, and recombines into the hole from which it left at time $t_r$. Figure 1B shows correlated recollision: the electron enters the continuum at time $t_b$, is accelerated by the driving field, interacts and excites the ion through the transfer of energy and momentum $-\vec{q}$ to a bound core-electron at time $t_{ee}$, and recombines into a different hole from which it left at time $t_r$. The phase of the matrix element describing the electron-electron interaction leads to significant modifications to the electron momentum (depicted by the change of color from blue to red), the time of recombination, and to an additional electric field-dependent phase $\Delta \Phi_{ee}$ imparted onto the recolliding electron during its trajectory \cite{14Patchkovskii_2012}.  Introducing a phased perturbing field in an \emph{in situ} measurement imparts a controlled phase on the recolliding electron that can be used to reconstruct the recolliding electron group delay through the relative phase $\phi$ between the driving and perturbing fields. This is shown by the dashed lines in Figs. 1A and B. For direct recollision, the continuum propagation dynamics are well understood and exhibit a linear group delay for short trajectories. Since recollision dynamics in the absence of multielectron resonances are well understood and structural effects do not affect continuum dynamics, deviations from this behavior can be attributed to multielectron interaction. Thus, by only focusing on continuum dynamics, \emph{in situ} techniques zero in on multielectron dynamics during a recollision trajectory.  

We experimentally study the giant plasmon resonance in Xe as observed through attosecond pulse generation. We then compare these measurements to a calculation of recollision-induced plasmonic excitation in C$_{60}$. C$_{60}$ is chosen for its similar resonance structure to Xe \cite{15PhysRevLett.94.065503} and computational accessibility.  Finally, we use the strong-field approximation (SFA) to describe the effect multielectron interactions have on semi-classical trajectories and \emph{in situ} measurement. 

\begin{figure}
	\includegraphics[width=\columnwidth]{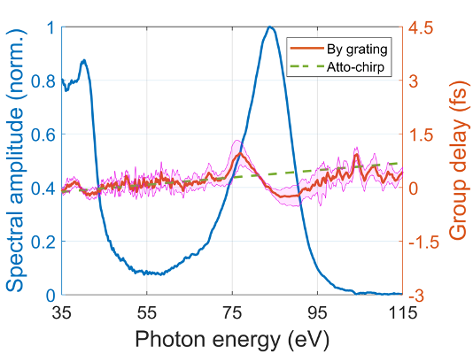}
	\caption{The spectral amplitude (blue) and group delay (orange) of an isolated attosecond pulse generated from Xe and reconstructed by a non-collinear single image \emph{in situ} measurement. The broad peak centered around 80 eV corresponds to the well-known giant plasmon resonance in the recollision spectrum of Xe. The measured group delay exhibits a quadratic variation from the linear atto-chirp (green dashed line) around the resonance. The standard deviation (27 data sets) of the group delay measurement outlines the plot in pink. Each group delay is shifted to synchronize the value of 39.4 eV.}	
\end{figure}

Figures 2 shows the experimentally acquired \emph{in situ} spectra measured from Xe. As described in the methods section, we use a few-cycle 1.8 \textmu m pulse with intensity of $2\times10^{14}$ W/cm$^2$.  We generate an attosecond pulse using polarization gating \cite{16Sola2006} and we use a version of \emph{in situ} measurement in which the temporal information of the generated attosecond pulse is encoded in the near-field spatial distribution of the generated spectrum through the variation of the spectral intensity with the vertical dimension. As described more fully in \cite{17ko2020nearfield} and in methods, this spatial mapping of the spectral intensity to emission time permits us to measure the recollision electron group delay. The measured group delay exhibits a linear chirp for energies below the plasmonic resonance in Xe. Near the resonance, however, we observe quadratic behavior – a clear deviation from the solely field driven direct-recollision electron continuum dynamics, shown by the dashed green line \cite{18doi:10.1080/09500340412331301542}. Such a large variation is in good agreement with theoretical results (14). The large change in group delay is due to the altered trajectory dynamics that depend on the driving electric field, the momentum transferred during the electron-electron interaction, the time of interaction, and the time of recombination. The nonlinear group delay indicates that the interaction of the recolliding electron with the remaining bound electrons is strong, despite the comparatively small region of interaction, causing the recollision electron trajectory to deviate strongly from that of the solely field-driven case. The shaded area indicates the standard deviation of the group delay measurement, for which we include 27 measurements for statistical analysis. The group delay measurements are synchronized at an energy of 39.4 eV.  

\begin{figure}
	\includegraphics[width=\columnwidth]{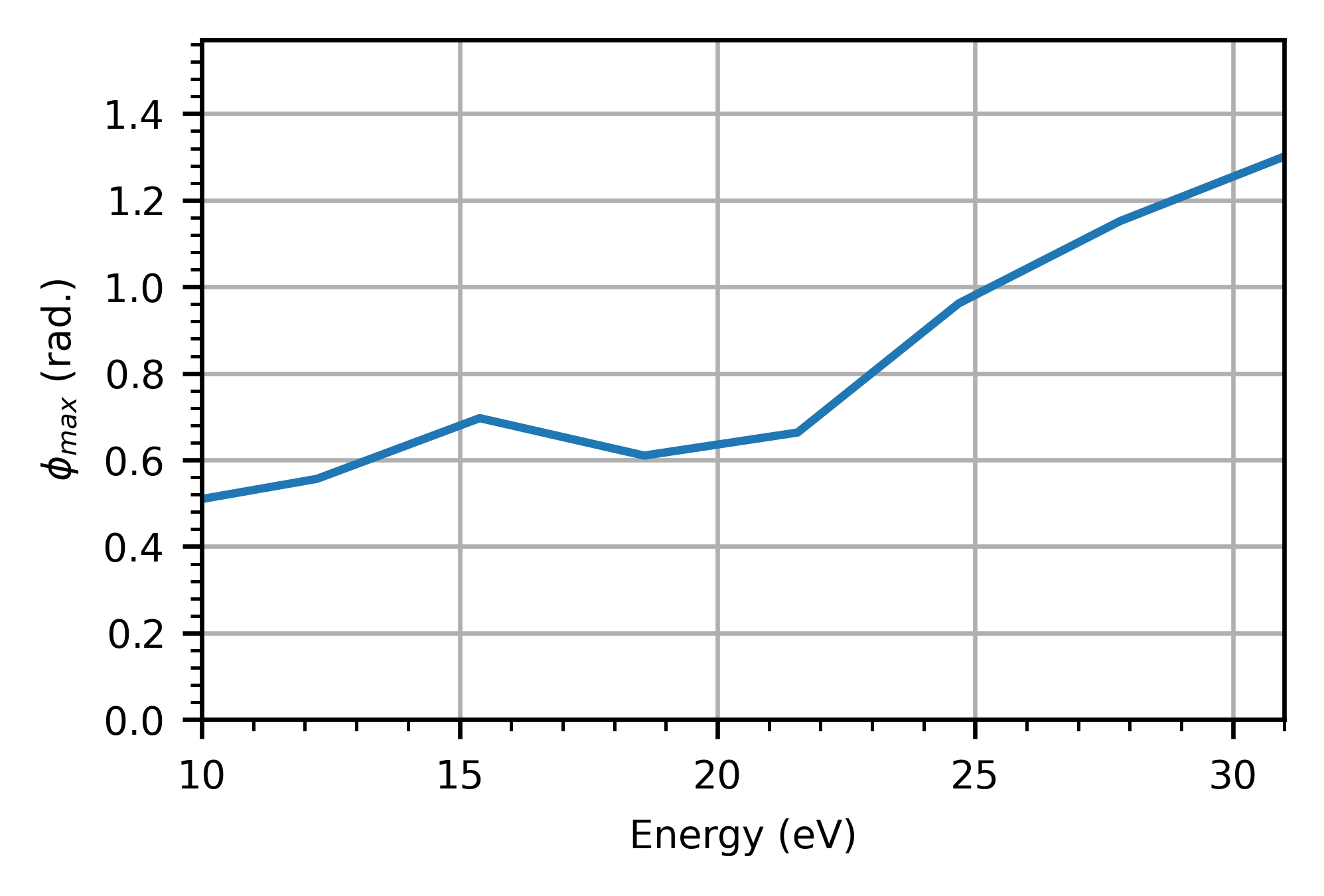}
	\caption{The relative phase between the fundamental and perturbing fields which maximizes the even-harmonic signal deviates from linear around the plasmon resonance between 15 and 22 eV. }	
\end{figure}

 To simulate the interaction between a recollision electron and a plasmon, we turn to time-dependent density functional theory (TD-DFT) \cite{19PhysRevLett.77.3865, 20PhysRevA.93.022506}.  We choose C$_{60}$, a system that exhibits a surface plasmon resonance near 20 eV \cite{14Patchkovskii_2012}, due to its computational accessibility.  To keep the dimensionality of the simulation tractable, we approximate C$_{60}$ as a spherical jellium shell (see Supplementary Information for more details), permitting azimuthal symmetry and a closed-shell electronic structure. We simulate a collinear $\omega-2\omega$ \emph{in situ} measurement with a linearly polarized 25 fs trapezoidal pulse of wavelength 800 nm pulse with intensity of $8\times10^{13}$ W/cm$^2$ and a co-propagating perturbing second-harmonic beam field with an intensity of $8\times10^9$ W/cm$^2$. The results are shown in Figure 3, where the relative phase between the fundamental and second harmonic fields that maximizes the even-harmonic signal is plotted against photon energy. For recollision in the absence of multielectron interaction, the expected result would vary linearly with energy and could be directly mapped to the recombination time. Around the surface plasmon between energies 16 and 22 eV, however, the behavior of the maximizing relative phase differs from linear due to the altered trajectory dynamics \cite{03Dudovich2006}. The qualitative agreement between the experimental measurement of Xe and simulation for C$_{60}$ shows that exciting the plasmon resonance during propagation strongly affects trajectory dynamics and that this effect can be measured using \emph{in situ} techniques.
 
Finally, we turn to the strong field approximation (SFA), the most intuitive model of electron recollision \cite{14Patchkovskii_2012, 21PhysRevA.49.2117}. In the SFA description of direct recollision, the electron is driven by the strong laser field after ionization at time $t_b$ and the attosecond pulse emission is determined by those trajectories that return to the parent ion at time $t_r$.  The solutions are determined solely by the force of the laser field acting on the electron in the continuum. Implemented with great success, it has been shown that only small corrections to the SFA are needed to account for the ionic Coulomb potential \cite{22Torlina_2017}. Our results (and the previous calculations for Xe \cite{02PhysRevLett.111.233005}) indicate that multielectron effects represent significant modification to the trajectory dynamics.

\begin{figure}
	\includegraphics[width=\columnwidth]{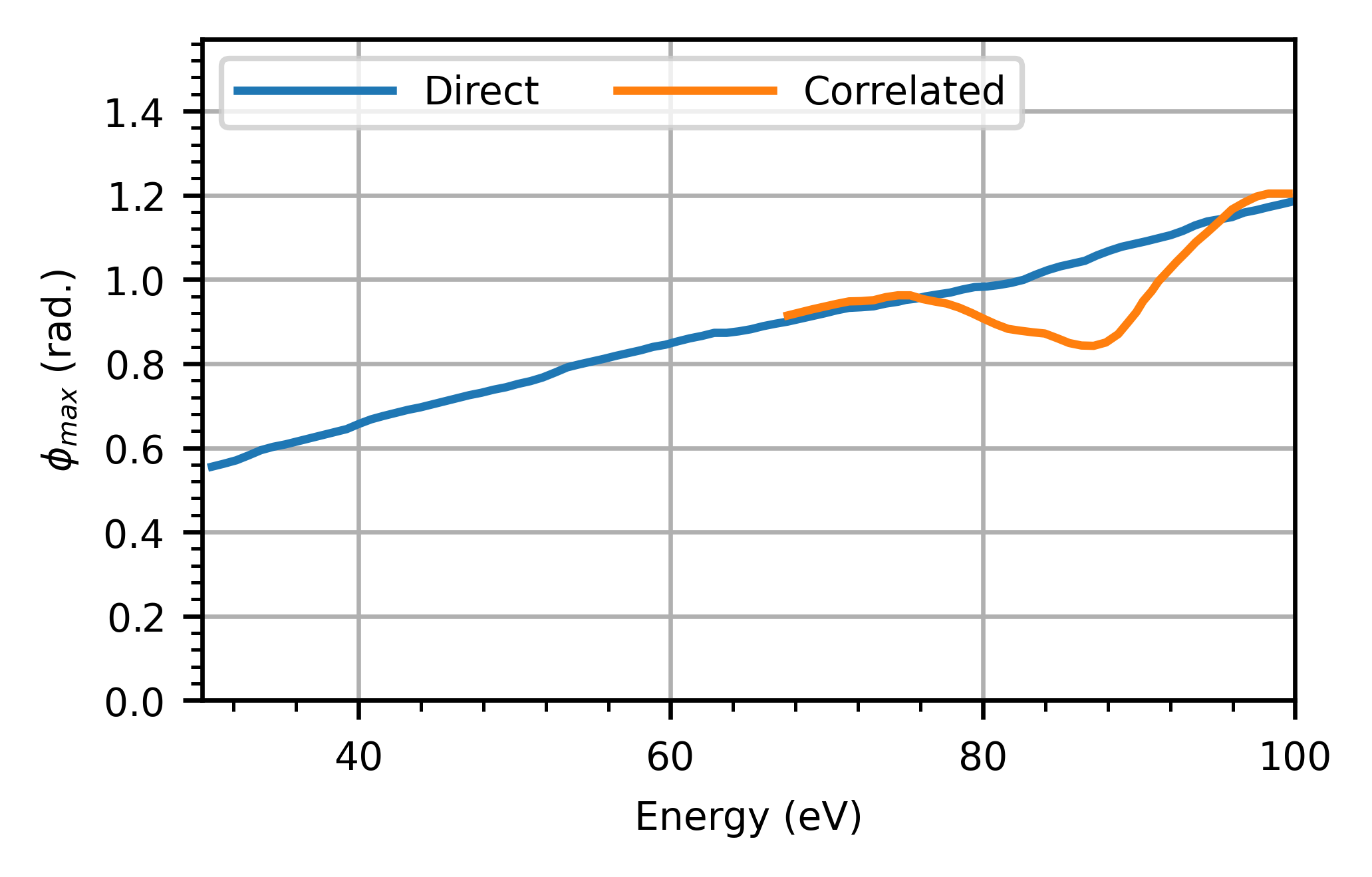}
	\caption{The simulated results of a collinear $\omega-2\omega$ \emph{in situ} measurement in Xe calculated using the correlated strong-field approximation (C-SFA) for a 1.8 \textmu m driving laser of intensity $10^{14}$ W/cm$^2$. The phase which maximizes the even-harmonic signal is shown in blue for the direct channel and red for the correlated channel. The direct channel exhibits the expected linear chirp, whereas the correlated channel deviates near the plasmon resonance.}	
\end{figure}
  
Following the approach for the correlated strong-field approximation (C-SFA) , we represent the multielectron interaction as a scattering process which modifies the recolliding electron momentum and trajectory.  In the C-SFA description of correlated recollision, the electron enters the continuum at time $t_b$ with momentum  $\vec{p}-\vec{q}$. At time $t_{ee}$, the recolliding electron transfers momentum $-\vec{q}$ to the ion, exciting it from a state of energy $I_{p,d}$ to a state of energy $I_{p,c}$. The recolliding electron then recombines with the ion at time $t_r$ with momentum  $\vec{p}$. In our model, the phase of the electron-electron interaction is assumed to only depend on the momentum transferred to the ion. For a constant electron-electron interaction phase, the solutions for the correlated channel are indistinguishable from the direct channel and the times of recombination and interaction are equal to the direct channel recombination time. For a varying electron-electron interaction phase, the times of recombination and interaction and the transferred momentum are modified, leading to a change in the recolliding electron phase, differentiating the correlated channel dynamics from those of the direct. In Fig. 4, we show the results of a simulated collinear $\omega-2\omega$ \emph{in situ} measurement \cite{03Dudovich2006} in Xe calculated using the C-SFA. In our calculations, $I_{p,d}$, the ionization of the potential before the electron-electron interaction, and $I_{p,c}$, the ionization potential of the atom after the electron-electron interaction, are equal to 12.2 eV, $I_{p,c}$ is 65 eV, respectively. The electron-electron interaction phase is modelled with a simple Lorentzian profile with a bandwidth of 20 eV centered around 90 eV. 

One advantage of the C-SFA treatment is that the separate channels leading to attosecond pulse emission can be deconvolved. Accordingly, we separately show the relative phase between the driving and perturbing fields which maximizes the even-harmonic signal for both the direct channel (blue) and the correlated channel (red). The direct channel exhibits the expected linear behavior, whereas the correlated channel strongly deviates from this behavior between the energies of 75 to 100 eV. As the correlated channel dominates the recollision spectrum in this spectral range due to its comparatively larger cross-section, these results agree with the experimental results presented in Figure 2, thus validating the effect electron-electron collisions have recollision trajectory dynamics. 

Although our experimental measurements were performed in a thin, gas jet where the density was low enough that phase matching is assured \cite{23Yakovlev:07}, the basic approach that we have introduced is applicable to atomic, molecular, and solid-state systems. Solid-state systems are more complex than gas-phased atoms and molecules, but \emph{in situ} techniques offer a route to measure multielectron dynamics in the conduction and valance bands; any force acting on the electron during its propagation is within the domain of \emph{in situ} techniques. 

There are three important implications of our results. First, we have experimentally confirmed that the recollision electron wave packet is not independent of the system. Plasmonic excitation, or any other multielectron interaction occurring during propagation, can influence the recollision electron trajectory.  We have used the spectral phase of a recolliding electron to measure this change.  However, the electron trajectory could also be measured with a second color which streaks a scattered recollision electron.  It should be possible to measure other multielectron effects such as non-sequential double ionization \cite{06Li2020}, the increase of high harmonic generation efficiency with atomic number \cite{07PhysRevLett.96.223902}, and electron correlation in the vicinity of spectral minima \cite{24PhysRevLett.112.153001}. 

Second, measuring the phase deviation of electron trajectories due to the presence of multielectron dynamics is equivalent to measuring the effect of an effective multielectron potential on the recolliding electron. Although density functional theory (DFT) has exhibited tremendous success in describing electronic structure, expressions for the effective potentials representing multielectron dynamics remain the principal approximation of the DFT formalism. Incorporating time-dependence into these approximations for exchange and correlation is an area of research that is inadequately understood \cite{25doi:10.1063/1.1904586}. The precision and control available to attosecond science can be applied to the direct study of time-dependent electron correlation in a practical and powerful way.   

Third, by manipulating core states, XUV emission involving correlated channels can be controlled. For the case of C$_{60}$ in particular, vibrational excitation, introducing a time-dependent reshaping of the molecular geometry, can be used to control the spectral or spatial characteristics of the dipole emission and, therefore, the enhanced XUV emission due to the plasmonic resonance \cite{26PhysRevA.88.043419}. Specifically, we envision a controlled vibrational excitation across the driving laser beam using an additional pulse to redirect a generated plasmon-enhanced XUV beam.
  
Finally, Coulomb interactions occurring prior to recombination are innately related to many of the questions facing researchers today concerning electron correlation and exchange. Our results are important for any attosecond experiment measuring many-body dynamics in atomic, molecular, and solid-state systems, including measuring the universal time response of electronic matter in response to the removal of an electron \cite{27PhysRevLett.94.033901}. The simplicity of \emph{in situ} measurement, namely that it does not require a transform-limited pulse for a transform-limited measurement, coupled with the importance of probing multielectron dynamics on the attosecond timescale will have a major impact on the physical sciences.  

\nocite{*}
\bibliography{characterizingMultielectronDynamicsDuringRecollision}

\end{document}